# Increasing The Sensitivity of a Surface Plasmon Resonance Sensor using $Ti_3C_2$ Mxene

Rostyslav Terekhov[1], Zoya Eremenko[1, 2], Sergii Kulish[3]

[1]O.Ya. Usykov Institute of Radiophysics and Electronics of the National Academy of Sciences of Ukraine, Kharkiv, Ukraine,

[2]Leibniz Institute for Solid State and Materials Research, 01069 Dresden, Germany,

[3]National Aerospace University "Kharkiv Aviation Institute", Kharkiv, Ukraine.

ABSTRACT

Sensors based on the phenomenon of surface plasmon resonance have limited sensitivity, up to 123 degree/Refractive Index Unit, which restricts their applicability in detecting subtle changes gin biological media (around $10^{-5}$ RIU). To address this, a sensor dependent on the Kretschmann-Reather configuration was modified using $Ti_3C_2$ MXene material. The sensor was modeled using the finite element method (FEM), yielding the dependence of the wave reflection parameter on the angle of incidence. Founded on these results, the sensitivity of the sensor was calculated and demonstrated a 10% improvement compared to the conventional Kretschmann-Reather configuration. In addition, human biological real-world samples were modeled using refractive index data, and the error of the proposed approach was determined to be approximately 3%.

INTRODUCTION

Sensors based on the surface plasmon resonance (SPR) method are widely used in various fields of science. For example, they are used to measure hemoglobin levels in blood [1], diagnose infectious and viral diseases at early stages [2], monitor environmental conditions [3]. Although these sensors are widely used in research, their sensitivity, which is up to 120 degree/RIU (the sensitivity of an SPR sensor is the ratio of the shift in the resonance angle to the change in the refractive index of the medium under investigation, measured in degrees per refractive index unit (RIU)) (e.g., glucose molecules).depending on the configuration, remains insufficient for measuring small ($10^{-5}$ RIU) changes in the refractive index of the measured material. This study proposes enhancing the sensitivity of an SPR sensor built on the ReatherKretschmann-Reather configuration (a setup used to excite surface plasmon polaritons at a metal/dielectric interface by directing light at a specific angle) by incorporating 2D nanomaterials.

An SPR sensor is sensitive to changes in the refractive index of a substance, [4]since surface plasmon polaritons are very sensitive to changes in the metal-dielectric boundary. [4]. Surface plasmon polaritons are electromagnetic waves that propagate along the metal/dielectric boundary and are usually observed in the infrared range. The name "surface plasmon polariton" refers to the combination of charge motion in metal ("surface plasmon") with electromagnetic waves in air or dielectric ("polariton")[5]. However, the SPR sensor also has significant limitations, such as nonspecific molecular binding to the sensor surface.

MXenes are a class of two-dimensional inorganic compounds consisting of atomically thin layers of transition metal carbides, nitrides, or carbonitrides. Over the past 10 years, their unique properties have attracted increasing attention due to their potential in various fields, such as electrochemical capacitors, electromagnetic interference shielding, water purification, medical applications, and more. The chemical formula of MXenes is $M_{n+1}X_nT_x$ (n = 1–3), where M represents a transition metal (Ti, Sc, V, Cr, etc.), and X represents carbon and/or nitrogen. MXenes are produced by selective etching of the A-layer from the MAX phase (a class of thermoelectric materials with the formula $M_{n+1}AX_n$, combining metallic electrical conductivity with ceramic thermal stability, e.g., $Ti_3AlC_3$) [6]. This allows the bonds between the layers to be broken and 2D nanostructures with a layered structure to be obtained.

The molecular properties and dynamics of biological substances under measurement conditions must be considered. In particular, proteins tend to be subjected to nonspecific adsorption onto vessel walls or sensor surfaces due to weak electrostatic and van der Waals interactions. The use of porous and hydrophilic MXene-based coatings reduces this effect, since their surface terminations (–OH, –O, –F) form a hydration layer that hinders protein adhesion, while micro- and mesopores introduce steric restrictions for large biomolecules [7]. This issue can be addressed using a porous, hydrophilic material that prevents biological molecules from adsorbing to the sensor surface. A group of 2D MXene nanomaterials with a porous structure and specific physicochemical properties was selected as the coating material. In particular, the hydrophilicity of MXene surfaces [8] and their abundance of surface terminations (–O, –OH, –F) enable the formation of a stable hydration layer that reduces nonspecific molecular adsorption. In addition, the surface chemistry of MXenes allows functionalization with bioligands, which improves selectivity and suppresses nonspecific binding. At the same time, their high electrical conductivity [9] does not directly affect adsorption, but ensures efficient plasmon coupling with the metal layer, thereby enhancing the sensitivity of the SPR sensor.

Recently, Bhawsar and Prabhu [10] achieved a sensitivity value of 80 degree/RIU with a SPR sensor with 10 layers of graphene at a wavelength of 633 nm. Bhisma Karki et al. [11] achieved a sensor sensitivity value of 123 degree/RIU in 2023 with a SPR sensor with a single layer of MXene and ZnO. The paper proposes a two-layer SPR sensor structure based on the Kretschmann-Reather configuration, consisting of a silver layer and a $Ti_3C_2$ MXene layer, operating at a wavelength of 633 nm. The article is structured as follows. The Introduction provides the motivation for this study. Section 2 explains the computer modeling used to obtain the results for different sensor structures. Section 3 presents the modeling results and their interpretation.

## BILAYER SPR SENSOR STUDY METHODS AND MATERIALS

Problem statement

The aim of this study is to theoretically model and investigate a sensor constructed on hybrid $Ti_3C_2$ MXene and silver nanostructures, with the goal of assessing the potential for enhancing sensitivity in measuring the properties of biological substances using the surface plasmon resonance (SPR) method in the Kretschmann-Reather configuration.

The objectives of this study include the following:

− To analyze the dependence of sensor performance on its geometry and material composition; grounded on data obtained from computer modeling of the sensor, investigate the influence of the sensor's geometry and materials on its sensitivity;

− To investigate the variation in wave reflection with respect to the angle of incidence for different refractive indices; simulate the wave reflection parameter in the sensor structure depending on the angle of incidence for different refractive indexes;To evaluate the potential of the biosensor for analyzing real biological samples.

− simulate the operation of the sensor with real-world biological samples using data on the refractive indexes of biological samples from the human body.

This study examines a two-layer SPR sensor structure (Fig. 1), consisting of a 55 nm silver layer and a 2.79 nm $Ti_3C_2$ MXene layer, using a probing light wavelength of 633 nm.

Structure of a surface plasmon resonance sensor based on MXene nanomaterial

The proposed sensor structure (Fig. 1) is constructed on the Kretschmann configuration [12] for SPR sensors, which employs a glass prism with a thin silver film. This well-established configuration is widely used because it provides stable fabrication, direct optical coupling between prism and metal, and high sensitivity to changes in the medium above the metal layer. In our work, this classical design is further modified by introducing an additional MXene layer, which enhances the plasmonic response and allows further improvement of sensor sensitivity.

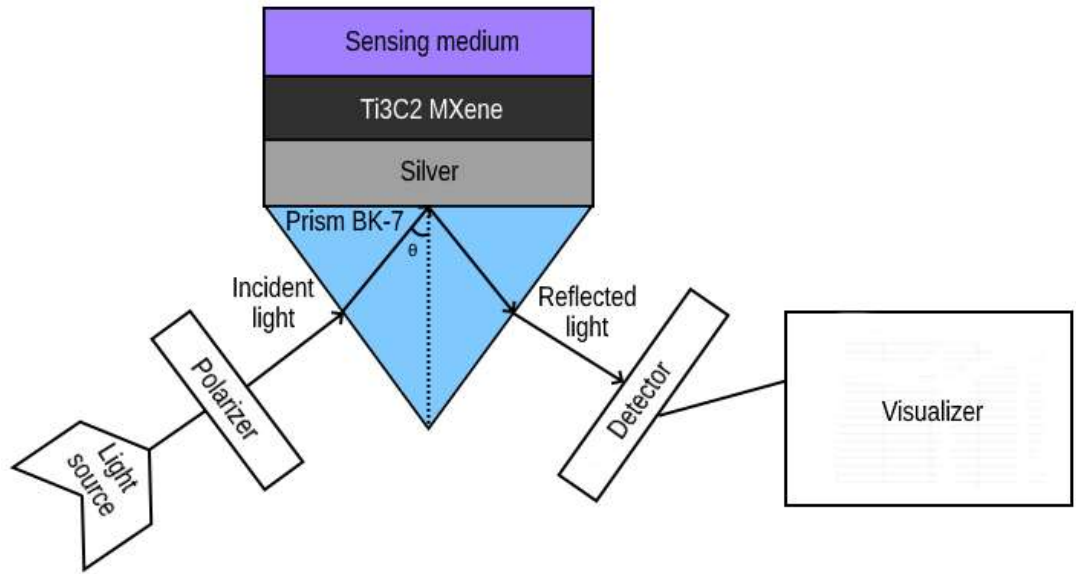

Figure 1 - Schematic diagram of the SPR sensor configuration used in this work, including a prism with Ag–MXene layers, a light source, and a detector.

Materials, model, and method of studying sensor properties

The Kretschmann configuration, modified with an MXene layer, was used as the sensor structure. It consists of a BK-7 glass prism, a thin silver film, a $Ti_3C_2$ MXene layer, and the biological sample under investigation. The prism serves to fix the angle of incidence of light. The sensor design was based on the Kretchmann configuration and was modified by introducing a layer of $Ti_3C_2$ MXene between the silver film and the biological sample. The complete design consisted of a BK-7 glass prism, a thin silver film, a layer of $Ti_3C_2$ MXene, and the biological sample under investigation. The MXene layer was introduced to improve the plasmonic characteristics: due to its high electrical conductivity, it improves the coupling of incident light with surface plasmons, and its hydrophilic surface and large number of terminal groups (–O, –OH, –F) reduce non-specific adsorption and stabilize the sensor interface. The real and imaginary parts of the refractive index for silver were taken from [13], and for MXene - from [14]. The studies were conducted using the finite element method (FEM).

Sensor parameters were modeled in two-dimensional space using COMSOL Multiphysics 6.2 (O.Ya. Usykov Institute for Radiophysics and Electronics, NAS of Ukraine (IRE) license No.

17078683). The Wave Optics module—specifically, the 'Electromagnetic Waves, Frequency Domain' interface—was employed to compute the harmonic distribution of the electromagnetic field in the frequency domain [15]. The wave equation was applied throughout the model. Periodic ports were defined to allow p-polarized light to be incident on the silver film at varying angles θ (from 0° to 89°) [16]. A geometric mesh was created, which is a discretization of the geometric model into elements used for the numerical solution of differential equations by the finite element method [17]. In COMSOL, the reflection coefficient was determined using the $S_{11}$.

## RESULTS

The paper investigates the parameters of the sensor depending on the geometry and materials of the structure to determine the maximum sensitivity and minimum reflection. The sensitivity calculated as:

$$S = \frac{\Delta\theta}{\Delta n}, \tag{1}$$

where, Δθ - is the shift in the resonance angle, measured in degrees; Δn - is the change in the refractive index of the sensing medium caused by the difference between two concentrations of the analyte under study.

The results are shown in Figure 2(a), which depicts the dependence of the minimum reflection parameter on the thickness of the silver film. As can be seen from the figure, values up to 20 nm do not meet the requirements for surface plasmon excitation, resulting in no surface plasmon resonance. The minimum reflection point is found at a silver film thickness of 55 nm. This is because resonance conditions arise when the wave vector of the incident light along the surface coincides with the wave vector of the surface plasmon. The figure shows that a thickness of 55 nm is optimal for achieving resonance, at which the energy of the incident light wave is maximally transferred to surface plasmons, and, accordingly, reflection is minimized. Figure 2 (b) shows the dependence of the reflection parameter on the angle of incidence of the wave, which has a minimum value at 67.5°. A sharp decrease in the reflection parameter indicates the presence of surface plasmon resonance when incident at this angle, called resonance angle. A change in the refractive index of the sample under study leads to a shift in the resonance angle (Fig. 2(b)), and the sensitivity value reaches 116 degree/RIU. Thus, the optimal thickness of the silver layer for our study is 55 nm.

A study was conducted on the effect of MXene layer thickness on the ReatherKretschmann-Reather structure, the results of which are shown in Figure 3 (a). The figure shows that an increase in thickness significantly affects the reflection parameter, starting from 15 nm, the minimum reflection is 0.9.

For further research, a 2.79 nm thick MXene layer will be used, which increases sensitivity, as shown in Figure 4(b), and the minimum reflection point reaches 0.5, which is sufficient for the resonance shift to be clearly visible.

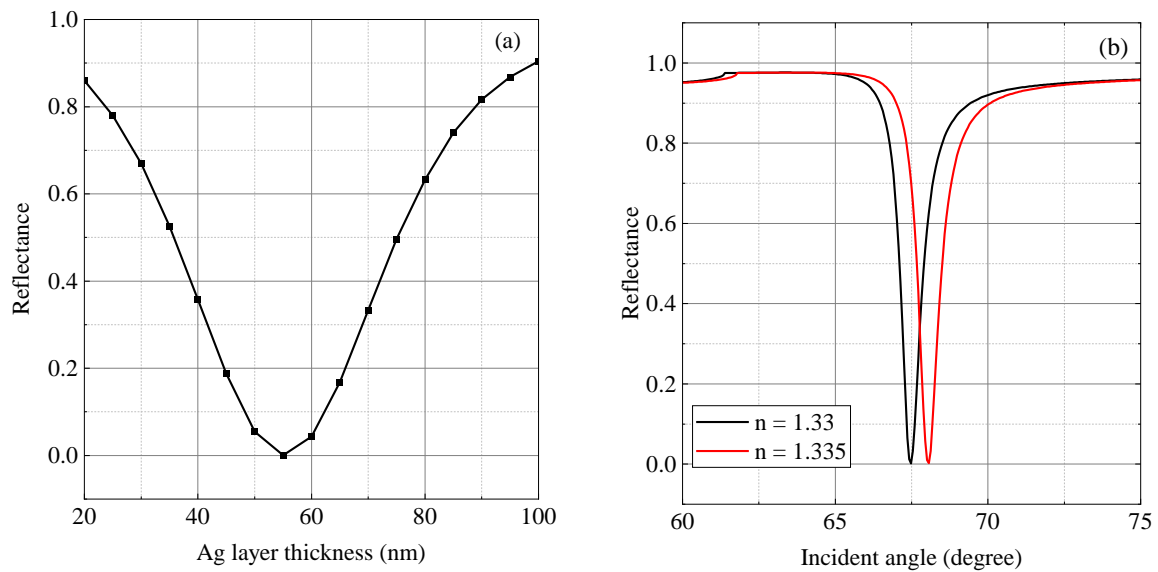

Figure 2. (a) Dependence of the minimum reflection point on the thickness of the silver layer. (b) Angular dependence of the reflection coefficient for the silver-based SPR sensor. The resonance shift was $\Delta\theta = 0.58°$, corresponding to a sensitivity of 116 degree/RIU.

A study was conducted on the effect of MXene layer thickness on the Kretschmann-Reather structure, the results of which are shown in Figure 3 (a). The figure shows that an increase in thickness significantly affects the reflection parameter, starting from 15 nm, the minimum reflection is 0.9.

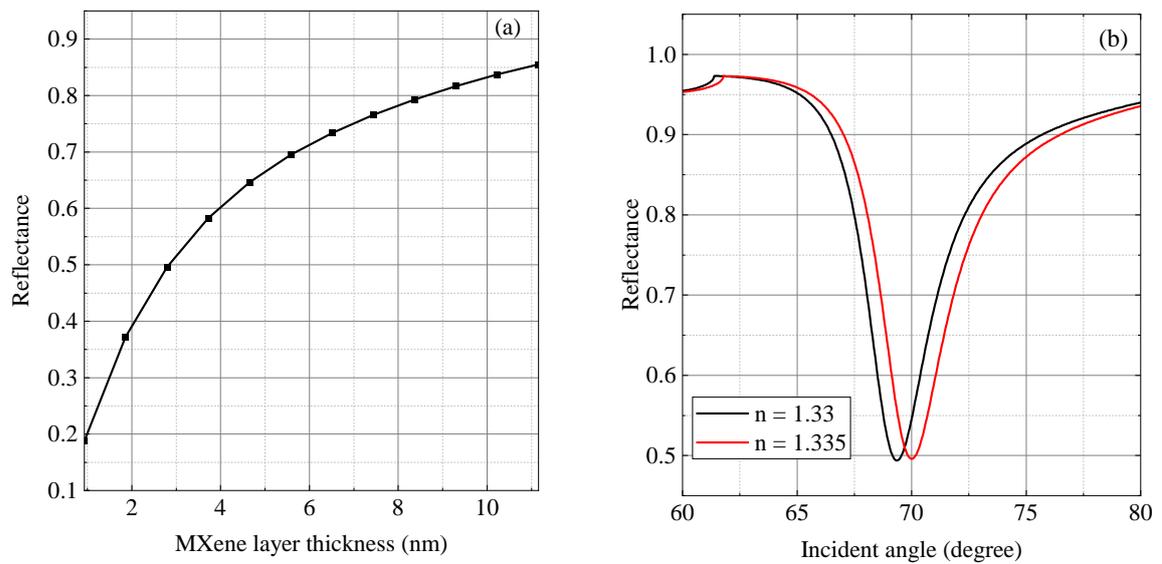

Figure 3. degree/RIU.(a) Dependence of the minimum reflection point on the thickness of the MXene layer at a fixed silver thickness of 55 nm. (b) Angular dependence of the reflection coefficient for the Ag–MXene SPR sensor. The resonance angle shift was Δθ = 0.638°, corresponding to a sensitivity of 127.6 degree/RIU.

For further research, a 2.79 nm thick MXene layer will be used, which increases sensitivity, as shown in Figure 4(b), and the minimum reflection point reaches 0.5, which is sufficient for the resonance shift to be clearly visible.

At the same time, there is an obvious increase in the angle delta between the analytes, which is a consequence of an increase in its refractive index. Using the sensitivity determination formula (3.1), the sensitivity of the SPR sensor structure with a silver layer and MXene was calculated. Thus, for a single-layer structure without an MXene layer, the sensitivity was 116 degree/RIU (Fig. 2 (b)), while for a double-layer structure, the sensitivity was 127.6 degree/RIU (Fig. 3 (b)). The increase in sensitivity was 10%. $Ti_3C_2$ MXene, as a two-dimensional material with high electrical conductivity, enhances the local electromagnetic field in SPR sensors. This is due to its metal-like behavior [18], which increases surface electron density and amplifies surface plasmon electromagnetic waves.

$Ti_3C_2$ actively absorbs light and generates photoinduced charge carriers. The combination of these properties with the SPR signal results in a significant enhancement of the electric field at the sensor interface [19].

To visualize the effect of MXene on surface plasmon resonance, the electric field intensity of the structure was modeled, and the results are shown in Figure 4.

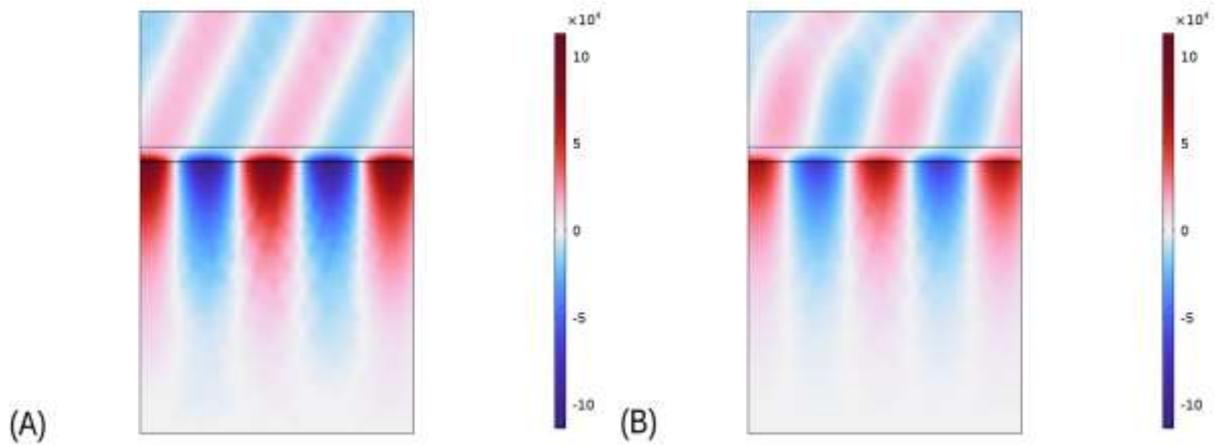

Figure 4. Image of the electric field strength of the sensor (V/m), where: a - field strength of the SPR sensor with a silver layer; b - field strength of the SPR sensor with a silver layer and MXene material.

The figure shows that the surface plasmon resonance intensity decreased when the MXene layer was applied. This can be explained by several factors. First, it is the high absorption of $Ti_3C_2$ MXene material, which has high dielectric permittivity and the ability to absorb electromagnetic waves in the terahertz range. This absorption causes the scattering of resonance energy, which leads to a decrease in the intensity of the localized field near the metal surface. Second, the introduction of an additional layer between the metal and the measuring medium changes the conditions for the occurrence of SPR. The MXene film creates a new interface that changes the local conditions for the propagation of plasmons and the wave vector of the surface plasmon. In addition, MXene is used for shielding due to its metallic nature. The thickness of $Ti_3C_2$ (~3 nm) is sufficient to partially shield the electromagnetic field, reducing the field near the metal-dielectric boundary. The thin film reduces the "flow" of the field into the medium below, and thus reduces the overall intensity of the SPR field.

A study was conducted on the use of a sensor based on the refractive index. The results obtained made it possible to determine the maximum value of the refractive index of the measured medium that the proposed sensor is capable of measuring. The figures show that at a value of 1.4, surface plasmons are not excited on the surface of the silver film. The reason is that surface plasmon polaritons cannot be excited when the refractive index (n) is too high, since the momentum of a free space photon is insufficient to match the higher momentum required for polaritons at this interface. This mismatch prevents effective coupling between the photon and the surface plasmon polaritons, making excitation impossible. When the refractive index of the sensitive medium exceeds a certain critical level, which has been found to be a refractive index of 1.4, no resonance is formed (Fig. 5). Therefore, the proposed sensor is capable of measuring the refractive index of a substance if it does not exceed 1.4. Figure 5 shows the values of the resonance angle, which, according to calculations, correspond to the refractive index of the material under study.

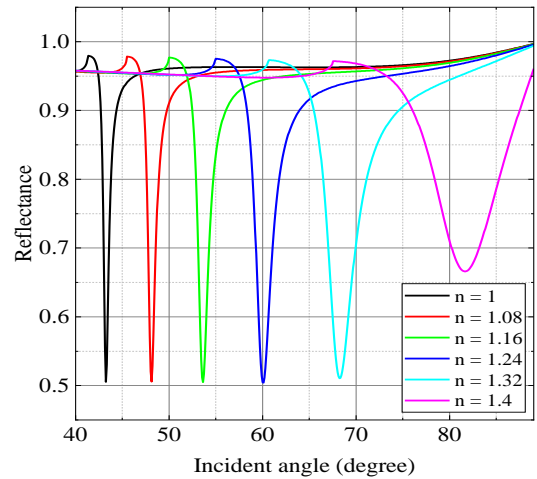

Figure 5. Dependence of the wave reflection parameter on the angle of incidence for different values of refractive index, the thickness of the silver layer is 55 nm, and the thickness of the MXene layer is 2.79 nm.

Given that biological material samples are stored in the form of aqueous solutions, this method can be effective in outpatient use. The concentration of the substance in the aqueous solution should also be taken into account. Obviously, the higher the concentration, the higher the refractive index, so this method can be used to determine the concentration of the active substance in solutions, whether it is glucose, hemoglobin, etc. The study examined real-world biological materials and determined the dependence of the resonance angle on the refractive index. It was established that the refractive index of the biological sample ranges from 1 to 1.4. The results of the study are shown in Figure 6.

The figure shows that the resonance angle increases with an increase in the refractive index of the substance being measured. The study was conducted using real-world biological materials, such as a sample of human liver tissue [20] and a sample of human colon tissue [21]. The results of the study showed a dependence of wave reflection on the angle of incidence, which is shown in Figure 7.

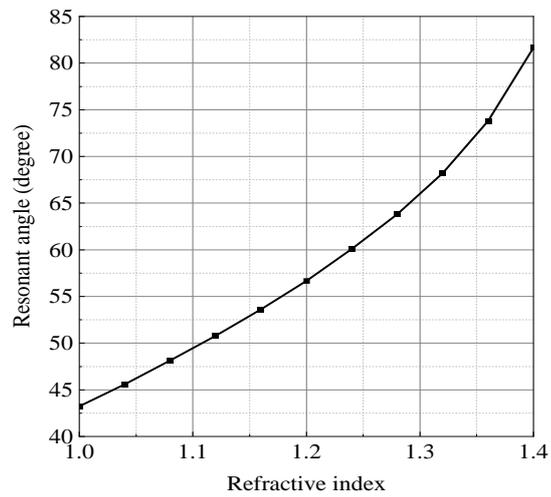

Figure 6. Dependence of the resonance angle on the refractive index of materials, the thickness of the silver layer is 55 nm, and the thickness of the MXene layer is 2.79 nm.

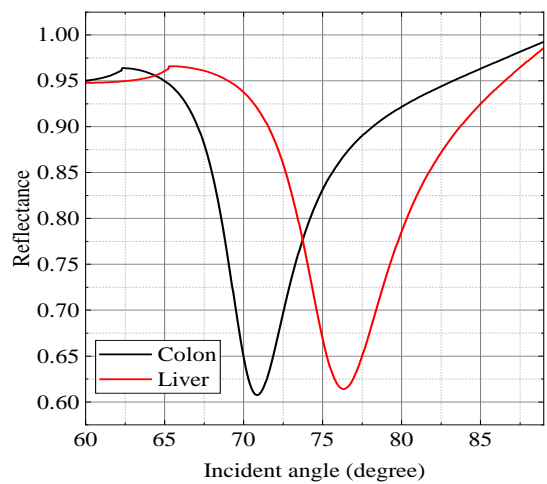

Figure 7. Dependence of wave reflection on its angle of incidence for real-world samples of human liver and colon. Wavelength (λ) is 633 nm, the thickness of the silver layer is 55 nm, and the thickness of the MXene layer is 2.79 nm.

The figure shows that the refractive index of a human colon sample at a wavelength of 633 nm is 1.341, with a resonance angle of 70.85°. The refractive index of the human liver sample is 1.3762, and the resonance angle is 76.3° [22]. To establish measurement accuracy, the relative measurement error was calculated as:

$$\delta = \frac{\Delta X}{X_{true}} \times 100\%$$

(2)

where, $\Delta X = X_{measured} - X_{true}$ – absolute error; $X_{true}$ – reference value; $X_{measured}$ – measured value;

Thus, based on the results of research, modeling, and calculations, it was established that the relative measurement error was approximately 3%. This error is due to the presence of an extinction coefficient in real-world biological samples, which affects the position of the resonance angle.

CONCLUSIONS

MXene–silver layers, using the finite element method implemented in COMSOL Multiphysics 6.2. The aim of this work is to investigate the potential for increasing the sensitivity of a surface plasmon resonance sensor built on the Kretschmann-Reather configuration by incorporating $Ti_3C_2$ MXene–silver layers. The analysis was carried out using the finite element method implemented in COMSOL Multiphysics 6.2.

It should be noted that all results were obtained by computer modeling; no physical sensor was fabricated at this stage. The results therefore demonstrate the theoretical potential of $Ti_3C_2$ MXene layers to improve SPR sensitivity, which can be further validated experimentally in future work.

To achieve this goal, the dependence of the sensor parameters on its geometry and material composition was investigated. The simulation results showed that the minimum reflection point is achieved at a silver thickness of 55 nm, which satisfies the resonance condition when the in-plane component of the incident light's wave vector matches that of the surface plasmon. The MXene layer thickness of 2.79 nm enabled the achievement of minimum reflection at the resonance point and a high-quality resonance factor. The initial sensitivity of the single-layer Kretschmann-Reather-based sensor was 116 degree/RIU, whereas the sensitivity increased to 127.6 degree/RIU in the double-layer structure incorporating $Ti_3C_2$ MXene. Studies have shown that changes in the geometry of the materials of the proposed two-layer structure resulted in a 10% increase in sensor sensitivity, attributed to the characteristic physical properties of MXene, in particular its high electrical conductivity and ability to support surface plasmon excitation at the nanomaterial interface. This improvement reduces the limitations of conventional SPR sensors and makes it possible to detect subtle changes in the refractive index of biological media, which was formulated as the aim of this work.

Studies showed that incorporating an MXene layer in the sensor structure reduced the electric field strength, which is due to the strong absorption of light in the terahertz range by MXene. In addition, the quality factor of the resonant structure was 20.

The measurement range of the proposed sensor was evaluated, which showed that the maximum refractive index that the sensor can measure is n = 1.4.

The sensor was tested with real-world biological samples, and it was shown that the proposed Sensor can be used for the study of biological samples. During the study, the relative measurement error was determined to be approximately 3%, which is due to the presence of an extinction coefficient in real-world biological samples that affects the position of the resonance angle.

ACKNOWLEDGMENT

This work was supported by grant № 0122U001687 of the NAS of Ukraine; Z. E. Eremenko acknowledges the funding from the European Union under the Marie Skłodowska-Curie grant agreement no. MSCA4Ukraine project number 1.4 - UKR - 1232611 - MSCA4Ukraine (IFW Dresden). 1.4 - UKR - 1232611 project has received funding through the MSCA4Ukraine project, which is funded by the European Union. Views and opinions expressed are however those of the author(s) only and do not necessarily reflect those of the European Union, the European Research Executive Agency or the MSCA4Ukraine Consortium. Neither the European Union nor the European Research Executive Agency, nor the MSCA4Ukraine Consortium as a whole nor any individual member institutions of the MSCA4Ukraine Consortium can be held responsible for them.